\documentclass{PoS}

\usepackage{amsmath,bbm,slashed,dsfont}
\usepackage{wrapfig}

\newcommand{\be}{\begin{equation}}
\newcommand{\ee}{\end{equation}}
\newcommand{\Z}{\mathcal{Z}}

\newcommand{\expv}[1]{\left \langle #1 \right \rangle}

\newcommand{\light}{{ud}}

\newcommand{\ml}{m_{ud}}
\newcommand{\pion}{{\pi^\pm}}
\newcommand{\Tr}{\textmd{Tr}}
\newcommand{\mut}{\mu_{I,pt}}
\newcommand{\Tt}{T_{pt}}
\newcommand{\Tc}{T_{pc}}

\title{QCD at nonzero isospin asymmetry\thanks{The research has been funded by the DFG via the
Emmy Noether Programme EN 1064/2-1 and SFB/TRR 55. B.B. has also received
support from the Frankfurter F\"orderverein f\"ur Physikalische
Grundlagenforschung.}}

\ShortTitle{QCD at nonzero isospin asymmetry}

\author{\speaker{Bastian B. Brandt}, Gergely Endr\H{o}di and Sebastian Schmalzbauer\\
        Institute for Theoretical Physics, Goethe University, Max-von-Laue-Strasse 1, 60438 Frankfurt am Main, Germany \\
        E-mail: \email{brandt@th.physik.uni-frankfurt.de}}

\abstract{We study the phase diagram and the thermodynamic properties of QCD at nonzero isospin asymmetry
at physical quark masses with staggered quarks. 
In particular, continuum results for the phase boundary between the normal and the pion
condensation phases and the chiral/deconfinement transition are presented. Our findings indicate
that the pion condensation phase is restricted to $T\lesssim170$~MeV for isospin chemical potentials up
to 325~MeV. We also use the data to test the range of validity of the Taylor expansion method and show
first results for the equation of state.}

\FullConference{XIII Quark Confinement and the Hadron Spectrum - Confinement2018\\
		31 July - 6 August 2018\\
		Maynooth University, Ireland}

\begin{document}

\section{Introduction}
\label{sec-0}

Most physical systems, such as nuclei, neutron stars, and possibly
the early Universe, feature an isospin asymmetry, i.e. an asymmetry between the number of
up ($u$) and down ($d$) quarks. In the grand canonical ensemble, QCD with two quark flavours
at finite density is described in terms of the independent isospin,
$\mu_I=(\mu_u-\mu_d)/2$, and baryon, $\mu_B=(\mu_u+\mu_d)/2$, chemical potentials.
The preferred tool to study QCD from first principles is Lattice QCD.
While most of the parameter space with $\mu_B\neq0$ suffers from the complex action problem,
QCD with {\it pure} isospin chemical potential, i.e., $\mu_I\neq0$ but $\mu_B=0$, has a real
and positive action
and is amenable to Monte-Carlo simulations. In most of the physical situations the effects due to
non-zero $\mu_B$ are expected to be dominant, but there are some cases where isospin might play the
major role, for instance in the early universe at large lepton
asymmetry~\cite{Wygas:2018otj} and for compact stars with pion
condensates~\cite{Migdal:1979je,Brandt:2018bwq}.
Furthermore, studying QCD at pure isospin chemical potential
is interesting in its own right. It
\begin{wrapfigure}{r}{7cm}
 \centering
 \vspace*{-3mm}
 \includegraphics[width=7cm]{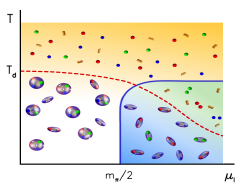}
 \caption{Conjectured phase diagram of QCD at pure isospin
 chemical potential (taken from~\cite{Brandt:2017oyy}).}
 \label{fig:phdiag}
\end{wrapfigure}
has a rich phase diagram, shown schematically in Fig.~\ref{fig:phdiag},
featuring a phase with Bose-Einstein condensation (BEC) of charged pions~\cite{Migdal:1978az,Ruck:1976zt}
and a hypothetical superconducting (BCS) phase at large $\mu_I$ on top of the standard hadronic and
quark-gluon plasma phases~\cite{Son:2000xc}.

In addition, QCD at pure isospin chemical potential shares a number of technical features with QCD at finite
baryon chemical potential, such as the Silverblaze phenomenon~\cite{Cohen:2003kd} and particle creation,
as well as a proliferation of low modes in the BEC phase. The latter demands the introduction of an infrared
regulator to facilitate simulations in the BEC phase~\cite{Kogut:2002tm,Kogut:2002zg,Endrodi:2014lja}.
A similar regulator might be necessary for simulations at non-zero $\mu_B$ beyond threshold.
QCD at pure isospin chemical potential is also the ideal test system for methods such as Taylor expansion,
which are commonly used to overcome the complex action problem for small $\mu_B$.

Following the initial studies of QCD at non-zero $\mu_I$ from
Refs.~\cite{Kogut:2002tm,Kogut:2002zg,Kogut:2004zg,deForcrand:2007uz,Detmold:2012wc,Endrodi:2014lja},
we have presented the first result for the continuum phase diagram with physical quark masses in
Ref.~\cite{Brandt:2017oyy} for $\mu_I\leq120$~MeV. Essential for this study has been the
introduction of a novel method for the extrapolation to vanishing regulator, $\lambda$, using the
singular values of the massive Dirac operator. We also compared our results at finite value of $\mu_I$
to results obtained from Taylor expansion~\cite{Brandt:2018omg}, where we also updated
the results for the BEC phase boundary to $\mu_I\leq325$~MeV. In this proceedings article
we summarise the studies mentioned above and show the updated phase diagram.
We also report on our ongoing measurements of the equation of state (EOS). First accounts of
our results have been presented in Refs.~\cite{Brandt:2016zdy,Brandt:2017zck}.
\nopagebreak
\section{Simulation setup and \boldmath $\lambda$-extrapolations}
\label{sec-1}

We study QCD at non-zero $\mu_I$ using $2+1$ dynamical quark flavours. The strange ($s$) quark,
has vanishing chemical potential, $\mu_s=0$, and we employ the fourth-root
trick. The fermion matrix of the light quark flavours includes a pionic source term
(see~\cite{Brandt:2017oyy}, for instance) as a regulator, with a prefactor $\lambda$ .
For the simulations we use a tree-level improved Symanzik gauge action and improve the fermion action by
using two steps of stout smearing. To tune the quark masses to their physical value we use the line of constant
physics from Ref.~\cite{Borsanyi:2010cj}. In our study we use four different temporal extents,
$N_t=6,\,8,\,10$ and $12$, corresponding to four different lattice spacings.

Our main observables to study the phase diagram are the renormalised pion and quark condensates, given by
(here $m_\light$ is the mass of the light quarks)
\be
\label{eq:renobs}
\Sigma_{\bar\psi\psi} = \frac{m_{\light}}{m_\pi^2 f_\pi^2} \left[ \expv{\bar\psi\psi}_{T,\mu_I} - \expv{\bar\psi\psi}_{0,0} \right] + 1, \quad\quad
\Sigma_{\pi} = \frac{m_{\light}}{m_\pi^2 f_\pi^2} \expv{\pion}\,,
\ee
where
\be
\expv{\pion}
= \frac{T}{V}\frac{\partial \log\Z}{\partial \lambda}, \quad\quad\quad
\expv{\bar\psi\psi} = \frac{T}{V}\frac{\partial \log\Z}{\partial \ml}\,.
\label{eq:pbpdef}
\ee
As an indicator for deconfinement we consider the renormalised Polyakov loop
\be
P_r (T,\mu_I) 
=  Z
\cdot P(T,\mu_I),
\quad\;
Z
= \left(\frac{P_\star}{P(T_\star,\mu_I=0)} \right)^{T_\star / T}, \quad\;
P = \expv{ \frac{1}{V}\! \sum_{V} \Tr \prod_{n_t=0}^{N_t-1} U_t(n) }
\label{eq:pren}
\ee
with $P_\star=P_r(T_\star,0)=1$ and $T_\star=162 \textmd{ MeV}$. Our main observable
to determine the EOS and the comparison to Taylor expansion is the isospin density
\be
\label{eq:nI-def}
\expv{n_I} = \frac{T}{V}\frac{\partial \log\Z}{\partial \mu_I}\,,
\ee
which does not require renormalisation.

The simulations are done at unphysical $\lambda>0$. Physical results are obtained in the
limit of vanishing regulator $\lambda\to0$. To this end we perform simulations at multiple
values of $\lambda$ and extrapolate the results to $\lambda=0$. These
extrapolations are the major challenge in the analysis, due to the pronounced
$\lambda$-dependence of most of the observables. To obtain
reliable $\lambda$-extrapolations over the whole parameter space and all observables
mentioned above, we have introduced an improvement program for the
$\lambda$-extrapolations~\cite{Brandt:2017oyy}. The program is based on the singular
value representation of the observables and consists of two steps: a ``valence
quark improvement'', corresponding to a reduction of the $\lambda$-dependence of the
observable, and an approximate reweighting to the $\lambda=0$ ensemble.
The remaining $\lambda$-extrapolation is mostly flat and can be carried out in a controlled
manner. From now on we will always work with $\lambda$-extrapolated observables.

\begin{figure}[t]
 \centering
 \includegraphics[width=7cm]{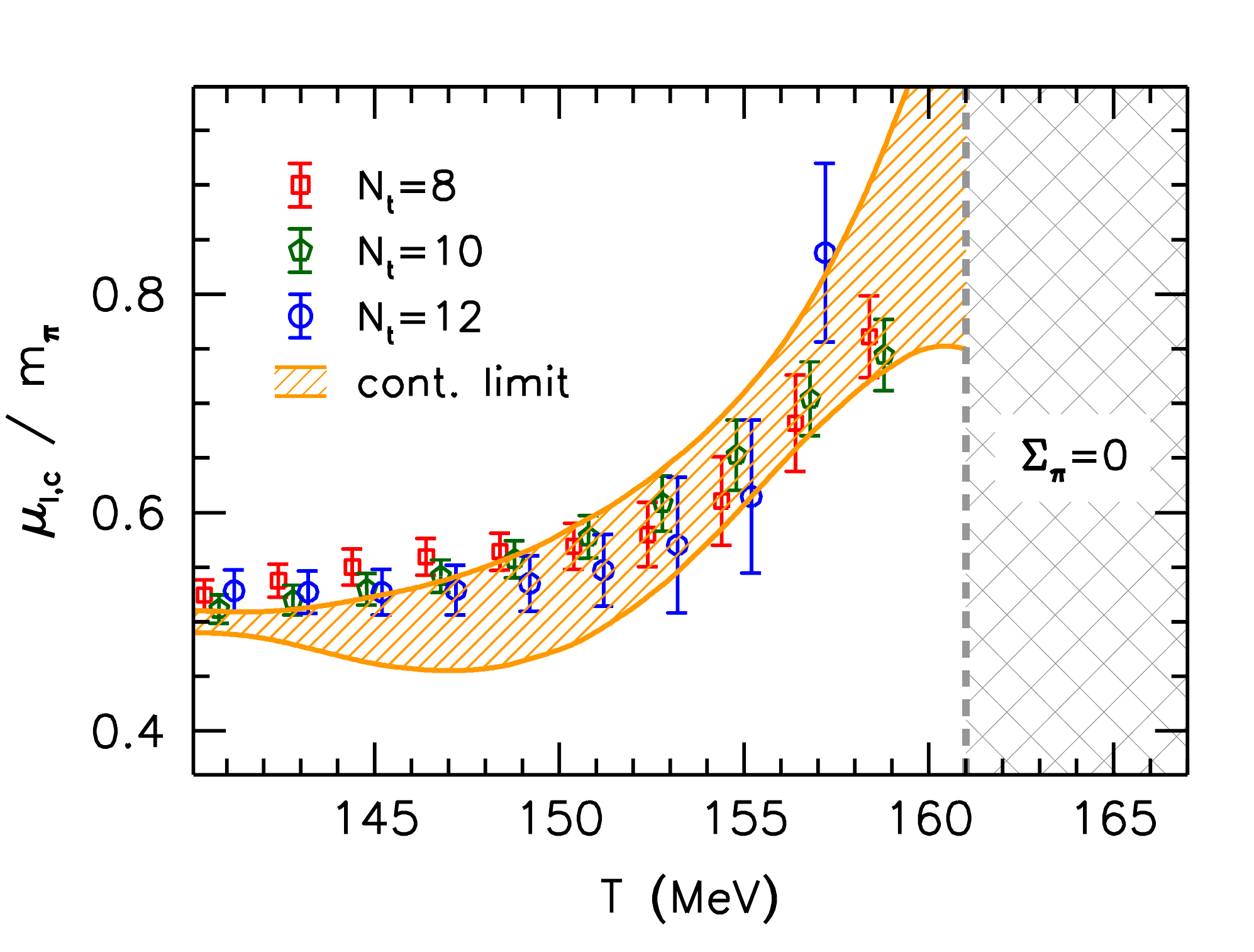}\hfill
 \includegraphics[width=7cm]{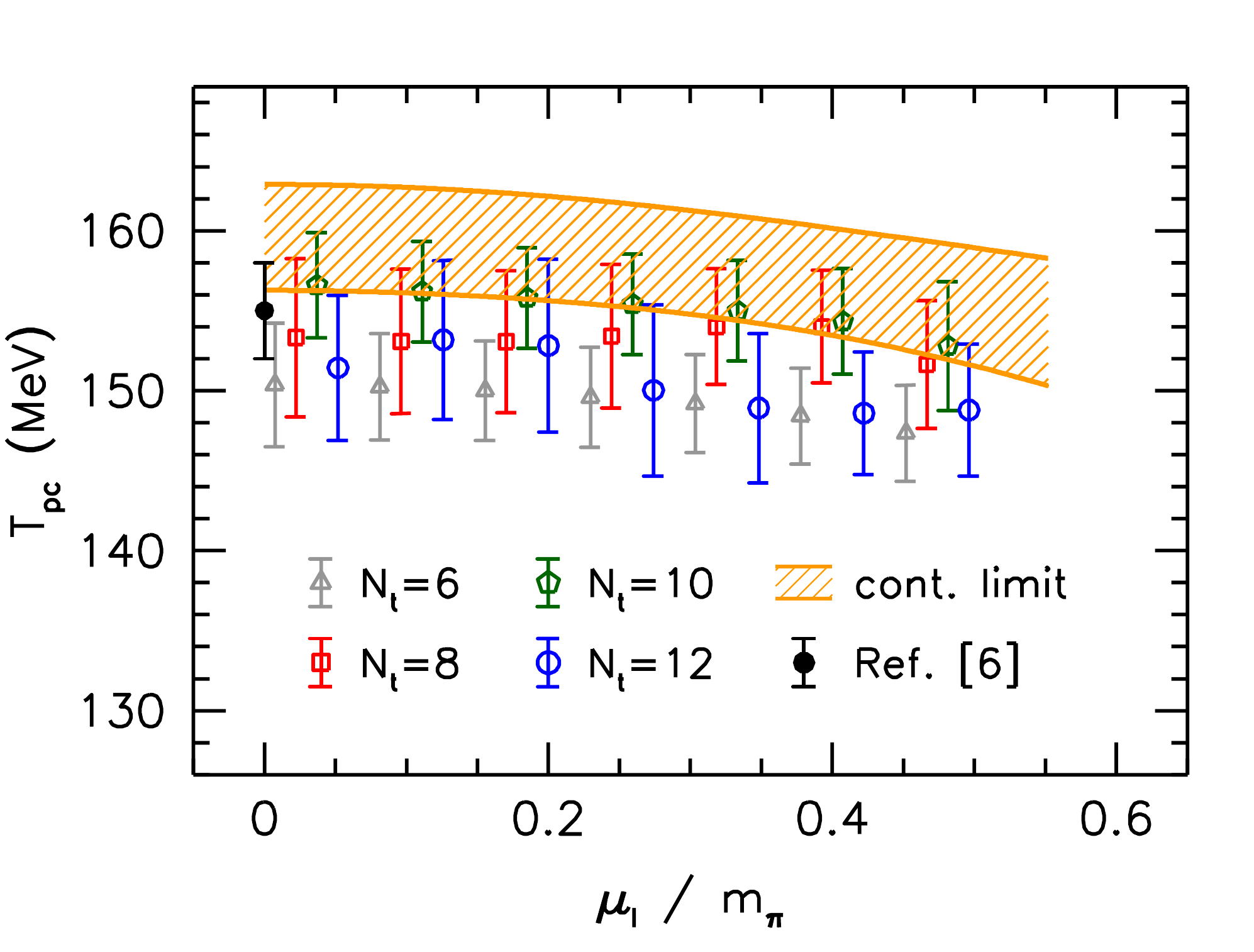}
 \caption{\label{fig:conti-extra}
 Continuum extrapolations for the BEC phase boundary (left panel) and the
 chiral crossover transition temperature (right panel). The yellow curves are the
 continuum extrapolations and the data points are the ones from the individual
 lattices which entered the fits. In the left panel, the shaded grey area
 represents the region where $\Sigma_\pi$ has been found to be consistent with
 zero within errors for $\mu_I\leq120$~MeV.
 }
\end{figure}

\section{Results for the phase diagram}
\label{sec-2}

\begin{figure}[t]
 \centering
 \includegraphics[width=7cm]{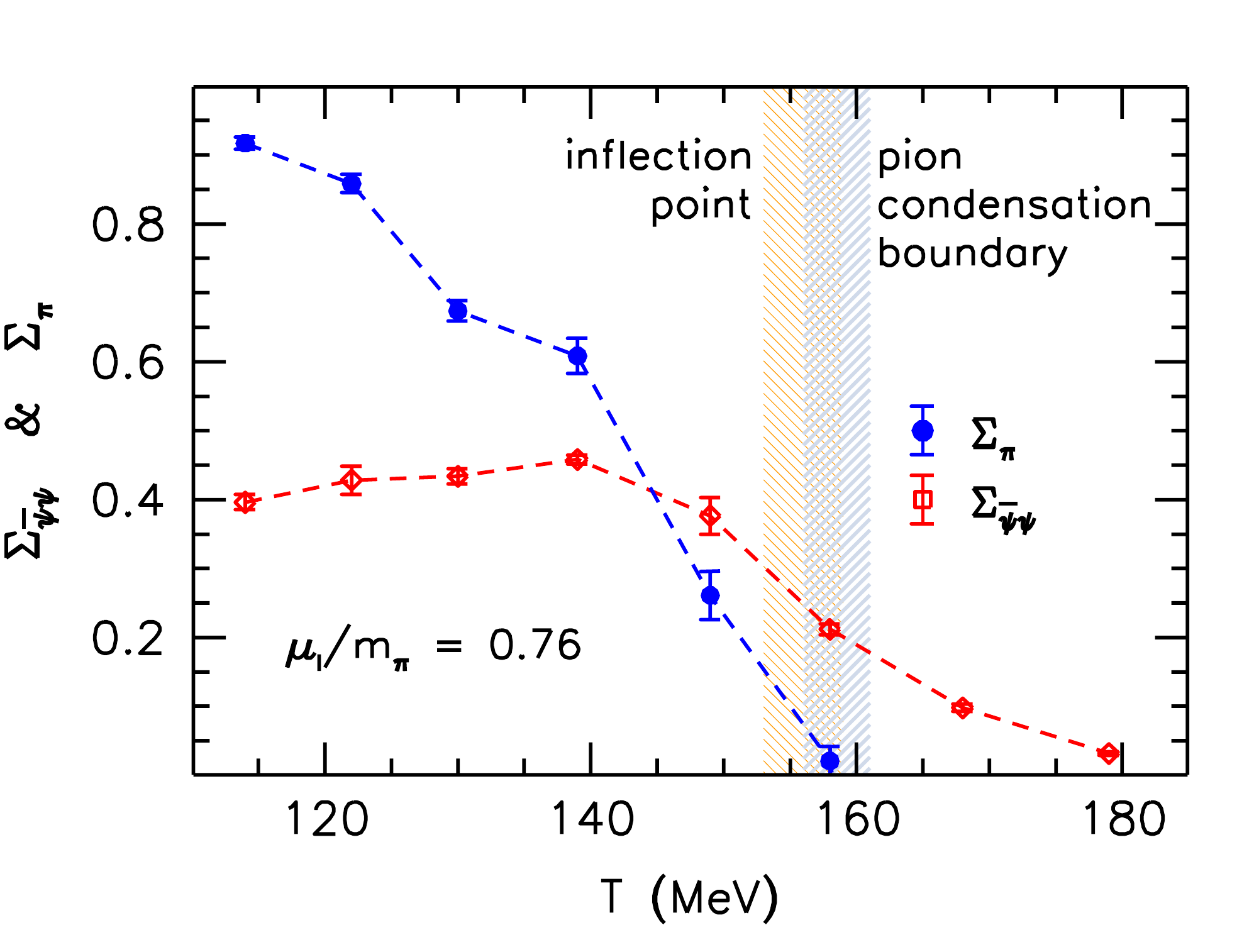} \hfill
 \includegraphics[width=7cm]{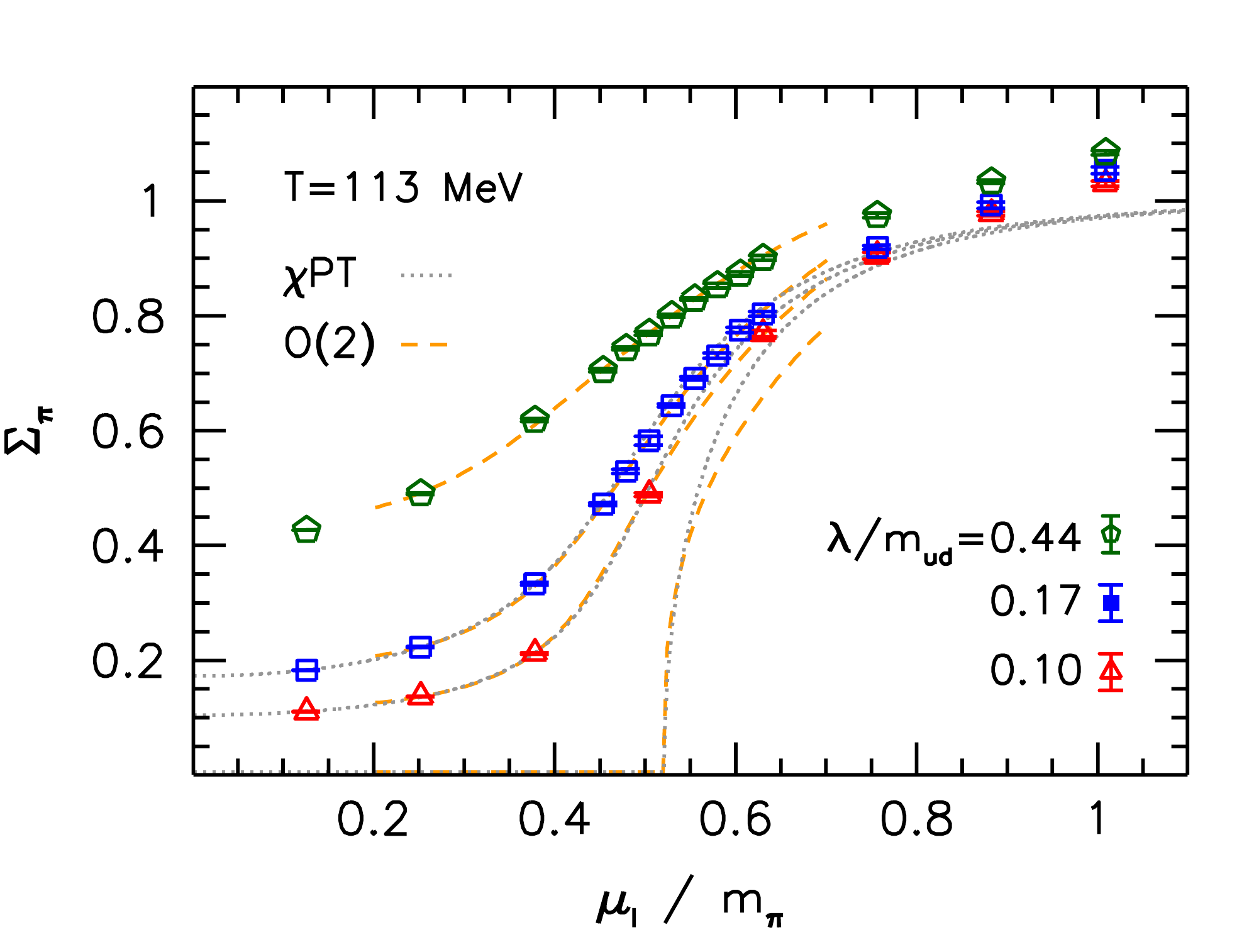}
 \caption{\label{fig:conti-extra-p2}
 Left panel: Pion and quark condensates as functions of $T$ for $\mu_I=103$~MeV on
 the $N_t=10$ ensembles. The light blue and orange areas mark the BEC phase
 boundary and the location of the inflection point of the condensate, respectively.
 Right panel: Comparison of the results for $\Sigma_\pi$ at nonzero $\lambda$ to $\chi$PT
 (dotted grey line) and to the critical behaviour of the $\mathrm{O}(2)$ universality class
 including scaling violations (dashed yellow line).
 }
\end{figure}

We start by presenting our results for the phase diagram at non-zero $\mu_I$.
The boundary of the BEC phase, $\mu_{I,c}(T)$, is determined by the points where
$\Sigma_\pi$ acquires a nonzero expectation value and the chiral crossover transition
temperature $\Tc(\mu_I)$ by the location of the inflection point of
$\Sigma_{\bar\psi\psi}$ with respect to $T$. To determine these phase boundaries,
we interpolate $\Sigma_\pi$ and $\Sigma_{\bar\psi\psi}$ for the individual lattices
using suitable two-dimensional spline fits where the nodepoints have been generated
via Monte-Carlo. For the continuum extrapolation, we parametrise the spline results
for $\mu_{I,c}(T)$ and $\Tc(\mu_I)$ by polynomials in $(T-T_0)$ (with $T_0=140$~MeV)
and $\mu_I^2$, respectively, including lattice spacing dependent coefficients. In
both cases we found the $N_t=6$ lattices to be outside of the scaling region. The
results for the continuum extrapolations are shown in Fig.~\ref{fig:conti-extra}. For
more details see Ref.~\cite{Brandt:2017oyy}. The two phase boundaries meet in a
pseudo-triple point at $\mu_I=\mut$ and $T=\Tt$ and are on top of each other from
that point on. This can be seen from the plot in the left panel of Fig.~\ref{fig:conti-extra-p2}.
The behaviour of $\Sigma_\pi$ and $\Sigma_{\bar\psi\psi}$ with $T$
for $\mu_I>\mut$ indicates that pion condensation and chiral symmetry
restoration occur at a similar temperature. A scaling analysis of $\Sigma_\pi$,
see the right panel of Fig.~\ref{fig:conti-extra-p2}, indicates that the transition
to the BEC phase is of 2nd order in the $\mathrm{O}(2)$ universality class,
as expected from the symmetry breaking pattern.

\begin{figure}[t]
 \centering
 \includegraphics[width=7cm]{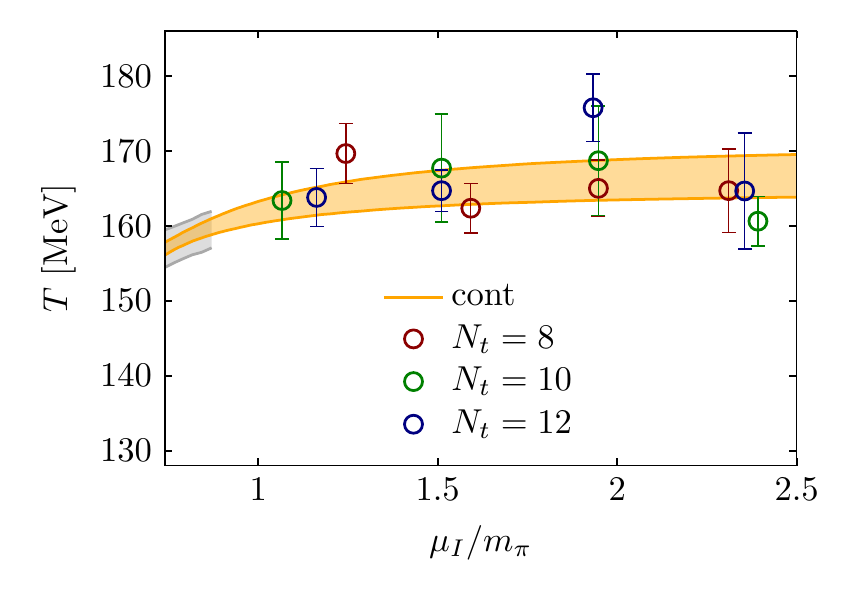} \hfill
 \includegraphics[width=7cm]{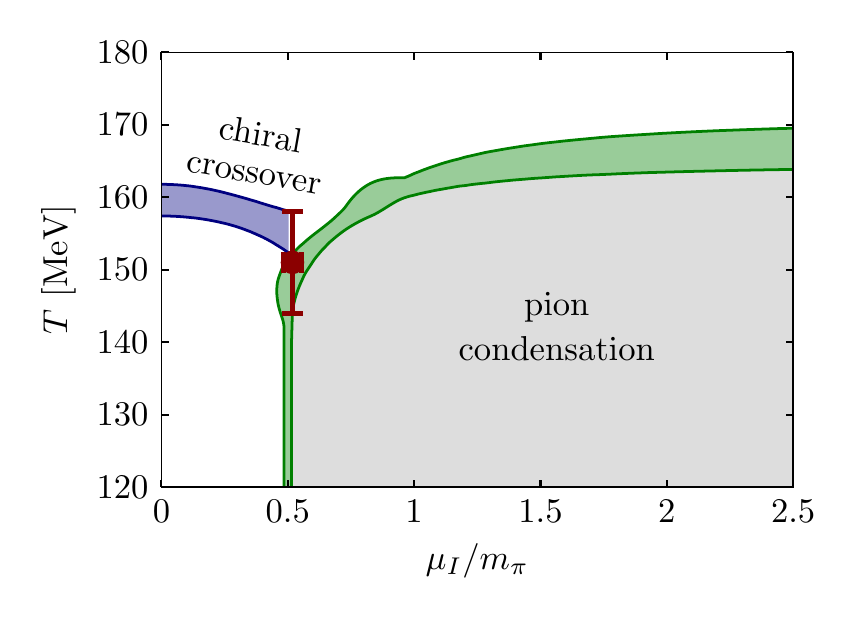}
 \caption{\label{fig:phdiag2}
 Left panel: Continuum extrapolation of the BEC phase boundary for $\mu_I>120$~MeV.
 The yellow curve is the continuum extrapolation, the data points are from the individual
 lattices and the grey band is the part of
 the continuum extrapolation for the BEC phase boundary for $\mu_I<120$~MeV,
 entering the fit for matching purpose.
 Right panel: QCD phase diagram for nonzero isospin chemical potential in the continuum limit. 
 Shown are the chiral crossover transition temperature $\Tc(\mu_I)$ (blue band) and the
 boundary $\mu_{I,c}(T)$ (green band) to the BEC phase (shaded grey area). The red point is the
 pseudo triple point, beyond which the two transitions coincide.
 }
\end{figure}

Recently we have also determined
the BEC phase boundary for $\mu_I>120$~MeV~\cite{Brandt:2018omg}, in this region
conveniently represented by a critical temperature $T_c(\mu_I)$. The continuum
limit has been performed by a fit to the form $d_1 + d_2/\mu_I^2$ with
$a^2$-dependent coefficients $d_1$ and $d_2$. In the fit we included data from
the continuum extrapolation for $\mu_I\leq120$~MeV from Fig.~\ref{fig:conti-extra}
for $T<161\textmd{ MeV}$ and 90~MeV~$\leq \mu_I \leq 120$~MeV, to smoothly connect
the two regions of the boundary. The resulting extrapolation is shown in
the left panel of Fig.~\ref{fig:phdiag2}. The updated continuum phase
diagram for $\mu_I\leq325$~MeV is shown in the right panel of Fig.~\ref{fig:phdiag2}.

Eventually, we are also interested in a possible crossover to the
BCS phase, which we expect to be related to the deconfinement transition in the BEC phase,
as indicated in Fig.~\ref{fig:phdiag}. First results on $N_t=6$ lattices have been
reported in Ref.~\cite{Brandt:2017oyy}.
A detailed study of the BCS phase and the deconfinement transition, however, demands
large values of $\mu_I$ at smaller temperatures, which we plan to study in the near future.

\section{A test for Taylor expansion}
\label{sec-3}

One of the possible methods to overcome the complex action problem
is the aforementioned Taylor expansion method. The key idea is to expand
observables around $\mu_B=0$, so that the resulting expressions include derivatives
at $\mu_B=0$ which can be computed in direct simulations. In practice, only a finite
number of expansion coefficients can be computed, so that the series has to be
truncated at that order. The main problem of the method is, that the reliability
region of the truncated series is unknown {\it a priori}. A similar expansion can
also be performed in $\mu_I$. For the isospin density, on which we will
focus from now on, the expansion takes the form
\be
\label{eq:nI-taylor}
\frac{\left\langle n_I \right\rangle
}{T^3} = c_2 \Big(\frac{\mu_I}{T}\Big) +
\frac{c_4}{6} \Big(\frac{\mu_I}{T}\Big)^3 +\ldots\,,
\ee
where $c_2$ and $c_4$ are the Taylor coefficients of the expansion of the pressure
in $\mu_I/T$ (see Ref.~\cite{Brandt:2018omg} for the details). The Taylor
coefficients for our action are available to us from Ref.~\cite{Borsanyi:2011sw}.

\begin{figure}[t]
 \centering
 \includegraphics[width=7cm]{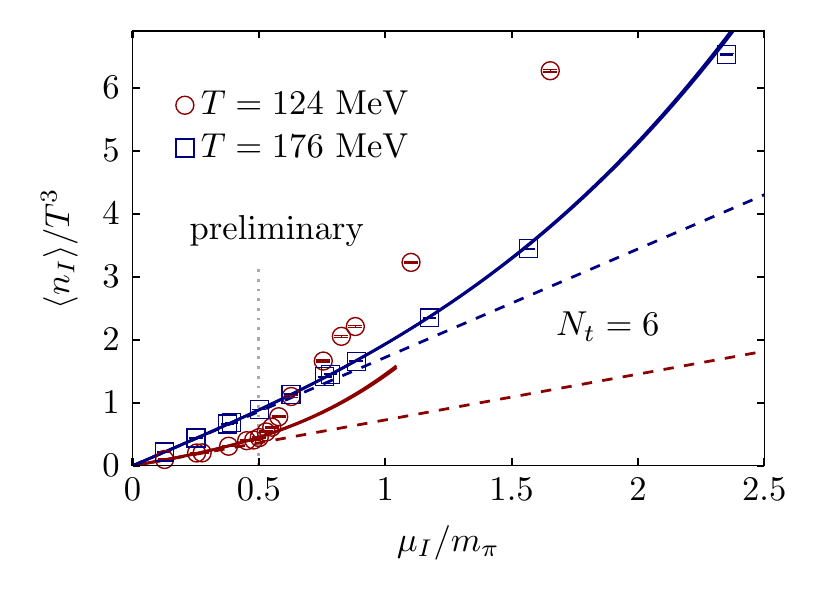} \hfill
 \includegraphics[width=7cm]{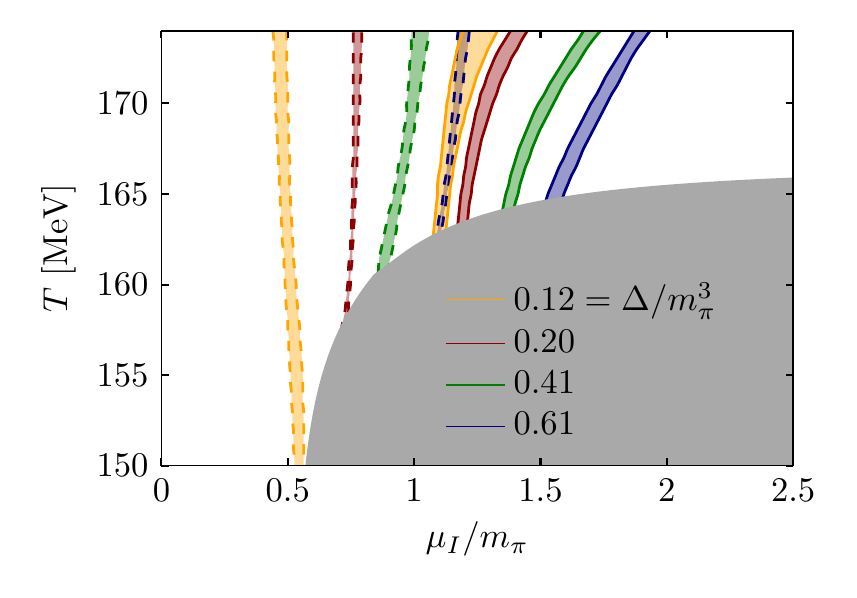}
 \caption{\label{fig:taylor}
 Left panel: Results for $\expv{n_I}$ at temperatures of 124 and 176~MeV on $N_t=6$
 lattices from direct simulations (red and blue points) in comparison
 to LO (dashed lines) and NLO (solid lines) Taylor expansions.
 Right panel: Contours of constant difference $\Delta^{\rm LO}$ (dashed bands)
 and $\Delta^{\rm NLO}$ (solid bands) for $N_t=8$.
 The shaded grey area represents the BEC phase.
 }
\end{figure}

In the left panel of Fig.~\ref{fig:taylor} we show the comparison between the full results
for $\expv{n_I}$ and the Taylor expansions to leading (LO) and next-to-leading (NLO) order
on the $N_t=6$ lattices
(all our data is reasonably far away from saturation with $\mu_Ia < 0.3$).
For $T=124$~MeV, the data reaches the BEC phase
boundary at $\mu_{I,c}\approx m_\pi/2$ and is in remarkable agreement with Taylor expansion
to LO and NLO up to this point. As expected, the Taylor expansion breaks down at the boundary,
showing large deviations to the lattice data. For $T=176$~MeV the data lies above the
BEC phase boundary and the agreement with Taylor expansion persists up to larger values
of $\mu_I$. At around $\mu_I/m_\pi\approx 0.6$ the data starts to favour the expansion to
NLO over the LO expansion. Higher orders become important at around $\mu_I/m_\pi\approx 1.6$.

For a quantitative comparison we consider contours with a constant
difference
\be
\label{eq:taylor-dif}
\Delta^{\rm LO/NLO} = \big\vert \expv{n_I} - \expv{n_I}^{\rm LO/NLO}
\big\vert
\ee
and focus on the high temperature region where the BEC phase
transition is absent. The contour lines are, once more, determined using two-dimensional
spline fits with Monte-Carlo generated nodepoints for $\Delta^{\rm LO/NLO}$. The contour
lines for different values of $\Delta$ at $N_t=8$ are shown in the right panel of
Fig.~\ref{fig:taylor}. One can clearly see the broader range of reliability of the NLO
expansion compared to the LO one and the tendency
of a better performance of Taylor expansion at larger temperatures.

\begin{figure}[t]
 \centering
 \includegraphics[width=7cm]{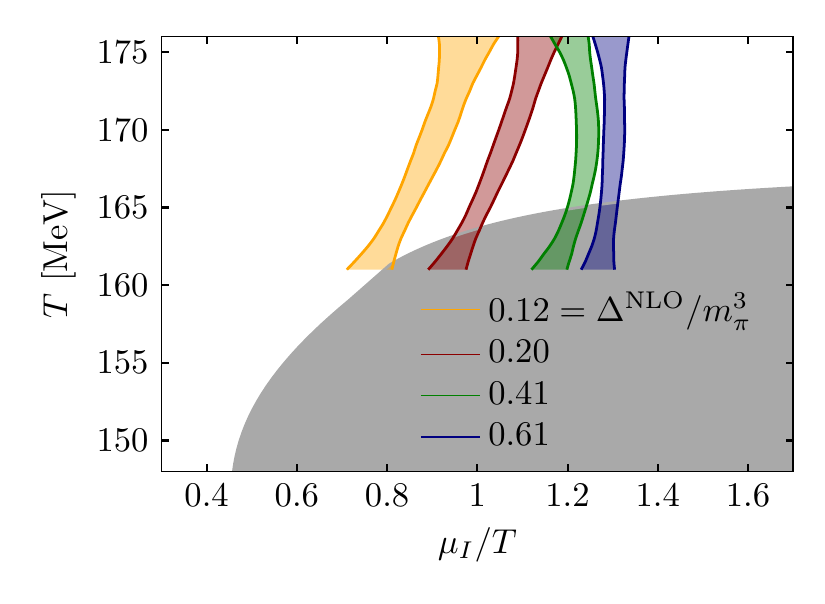} \hfill
 \includegraphics[width=7cm]{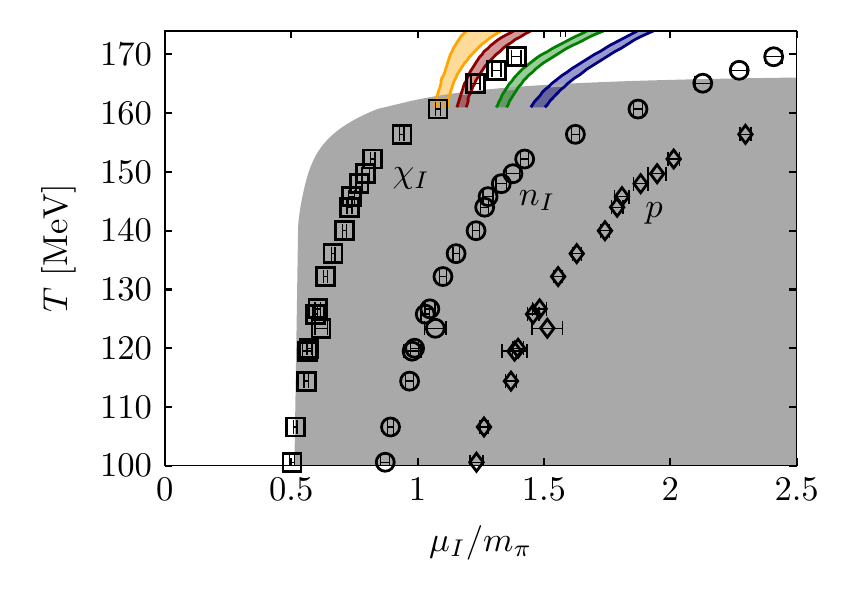}
 \caption{\label{fig:taylor-convr}
 Left panel: Continuum results for the contours of constant difference
 $\Delta^{\rm NLO}$. The grey shaded area indicates the BEC phase.
 Right panel: The leading-order estimators for the radius of convergence
 on our $N_t=8$ ensembles compared to the boundary of the BEC phase (grey area) 
 and the contours of $\Delta^{\rm NLO}$ (coloured bands).
 }
\end{figure}

To investigate the range of applicability of the NLO expansion in the continuum, we
extrapolate the contour lines using a second order polynomial in $(T-T_0)$ with
lattice spacing dependent coefficients and $T_0=140$~MeV. As before, including only
data with $N_t\geq8$. The continuum contour lines
are shown in the left panel of Fig.~\ref{fig:taylor-convr} versus $\mu_I/T$. As above we
observe deviations from the vertical $\mu/T=\textmd{const}.$ lines and the tendency
for a shift to larger values of $\mu_I/T$ with increasing temperature.

The Taylor expansion can also be used to test for the existence of a phase transition
at finite chemical potential, which should show up as a finite radius of
convergence of the series~\cite{Gavai:2004sd}. We test this method with the BEC phase
boundary. The radius of convergence $r$ for the Taylor series of $\expv{n_I}$ can be defined as
\be
\label{eq:conv-rad-nI}
r = \lim_{n\to\infty} r_n(n_I), \quad\frac{r_n(n_I)}{T} = \sqrt{ \frac{c_{n}}{c_{n+2}} (n+1)n } \,.
\ee
Similar estimators for $r$ can also be obtained from the series of the pressure $p$ and the susceptibility
$\expv{\chi_I}=\partial \expv{n_I}/\partial\mu_I$. While all estimates have to agree in the limit
$n\to\infty$, they differ at finite $n$. Currently we have access to the $n=2$ estimators only and, thus,
cannot investigate the $n\to\infty$ limit. In Fig.~\ref{fig:taylor-convr} (right panel)
we show the estimators $r_2$ for $N_t=8$ from the different observables. In the vicinity of the upper
BEC phase boundary, the estimators show a considerable change of slope,
indicating a possible agreement of the curves in the limit $n\to\infty$\footnote{Note, that for a general
singularity in the complex $\mu_I$-plane this limit is not guaranteed to exist -- see
Ref.~\cite{Vovchenko:2017gkg} for a counter example. We also ignored subtle issues regarding the estimators
of the radius of convergence at finite volumes, see Ref.~\cite{Stephanov:2006dn}, for instance.}.
It is interesting to note
that $r_2(\chi_I)$ is surprisingly close to the phase boundary for low temperatures,
while the other $r_2$ tend to overestimate $r$. This, likely accidental, agreement is consistent with findings
in a quark-meson model~\cite{Karsch:2010hm}, and in toy models of QCD with imaginary chemical
potentials~\cite{DElia:2016jqh}.

\section{Equation of State}
\label{sec-4}

For the study of nuclear- and astrophysical systems, knowledge about the EOS is of
fundamental importance. On top of the contribution from non-zero $\mu_B$, it will also receive
contributions from the isospin sector. Here we focus on the EOS at pure isospin
chemical potential. Knowing the pressure at vanishing $\mu_I$,
all thermodynamic quantities can be determined from the $n_I$. In particular,
the pressure can be written as
\be
\label{eq:pint}
 p(T,\mu_I) = p(T,0) + \int_0^{\mu_I} d\mu'_I \, n_I(T,\mu'_I)
 \equiv p(T,0) + \Delta p(T,\mu_I) \,.
\ee
Knowledge about $p$ and the trace anomaly \\
\begin{wrapfigure}{r}{7cm}
 \centering
 \vspace*{-3mm}
 \includegraphics[width=7cm]{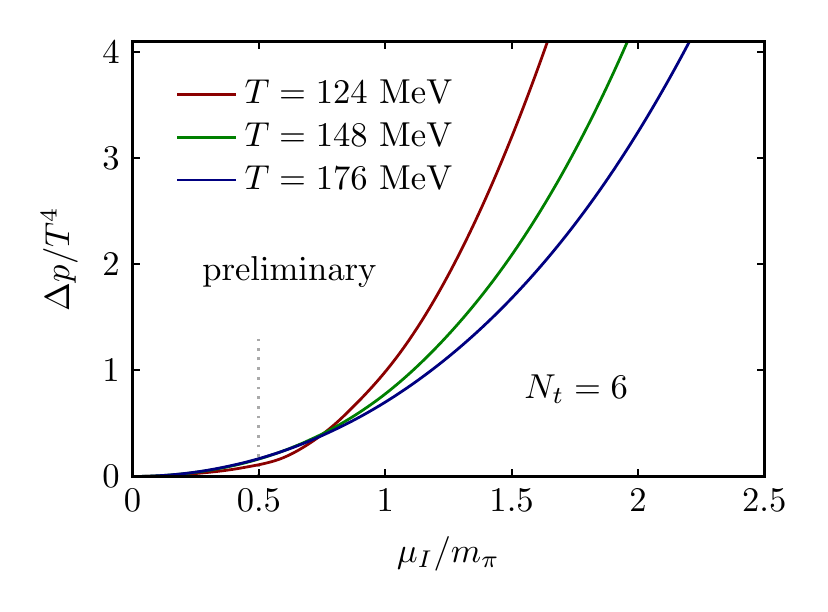}
 \caption{\label{fig:eos}
 Results for $\Delta p(T,\mu_I)$ for different temperature values at $N_t=6$.
 }
\end{wrapfigure}
\be
 \label{eq:tr-anom}
 \frac{I}{T^4} = \frac{\epsilon - 3p}{T^4} = T
 \frac{\partial}{\partial T} \frac{p}{T^4} + \frac{\mu_I n_I}{T^4} \,,
\ee
which can be extracted from $p$ and $n_I$, is sufficient for the computation of all of the
other thermodynamic quantities.

We evaluate $\Delta p(T,\mu_I)$ by integrating a cubic spline interpolation of the data
for $\left\langle n_I \right\rangle$. The results for $\Delta p(T,\mu_I)$ at different $T$
for $N_t=6$ are shown in Fig.~\ref{fig:eos}. Together with the results for the $\mu_I=0$
pressure from~\cite{Borsanyi:2010cj} these results give the full pressure. Using this approach
we calculated the EOS at $T=0$ in~\cite{Brandt:2018bwq}.


\vspace*{-3mm}

\begin{thebibliography}{99}
\bibitem{Wygas:2018otj}
  M.~M.~Wygas {\it et al},
  arXiv:1807.10815 [hep-ph].
\bibitem{Migdal:1979je}
  A.~B.~Migdal, A.~I.~Chernoutsan and I.~N.~Mishustin,
  Phys.\ Lett.\  {\bf 83B} (1979) 158.
\bibitem{Brandt:2018bwq}
  B.~B.~Brandt {\it et al},
  arXiv:1802.06685.
\bibitem{Migdal:1978az}
  A.~B.~Migdal,
  Rev.\ Mod.\ Phys.\  {\bf 50} (1978) 107.
\bibitem{Ruck:1976zt}
  V.~Ruck, M.~Gyulassy and W.~Greiner,
  Z.\ Phys.\ A {\bf 277} (1976) 391.
\bibitem{Son:2000xc}
  D.~T.~Son and M.~A.~Stephanov,
  Phys.\ Rev.\ Lett.\  {\bf 86} (2001) 592
  [hep-ph/0005225].
\bibitem{Cohen:2003kd}
  T.~D.~.Cohen,
  Phys.\ Rev.\ Lett.\  {\bf 91} (2003) 222001
  [hep-ph/0307089].
\bibitem{Kogut:2002tm}
  J.~B.~Kogut and D.~K.~Sinclair,
  Phys.\ Rev.\ D {\bf 66} (2002) 014508
  [hep-lat/0201017].
\bibitem{Kogut:2002zg}
  J.~B.~Kogut and D.~K.~Sinclair,
  Phys.\ Rev.\ D {\bf 66} (2002) 034505
  [hep-lat/0202028].
\bibitem{Endrodi:2014lja}
  G.~Endr\H{o}di,
  Phys.\ Rev.\ D {\bf 90} (2014) no.9,  094501
  [arXiv:1407.1216].
\bibitem{Kogut:2004zg}
  J.~B.~Kogut and D.~K.~Sinclair,
  Phys.\ Rev.\ D {\bf 70} (2004) 094501
  [hep-lat/0407027].
\bibitem{deForcrand:2007uz}
  P.~de Forcrand, M.~A.~Stephanov and U.~Wenger,
  PoS LATTICE {\bf 2007} (2007) 237
  [arXiv:0711.0023].
\bibitem{Detmold:2012wc}
  W.~Detmold, K.~Orginos and Z.~Shi,
  Phys.\ Rev.\ D {\bf 86} (2012) 054507
  [arXiv:1205.4224].
\bibitem{Brandt:2017oyy}
  B.~B.~Brandt {\it et al},
  Phys.\ Rev.\ D {\bf 97} (2018) no.5,  054514
  [arXiv:1712.08190].
\bibitem{Brandt:2018omg}
  B.~B.~Brandt and G.~Endr\H{o}di,
  arXiv:1810.11045.
\bibitem{Brandt:2016zdy}
  B.~B.~Brandt and G.~Endr\H{o}di,
  PoS LATTICE {\bf 2016} (2016) 039
  [arXiv:1611.06758].
\bibitem{Brandt:2017zck}
  B.~B.~Brandt, G.~Endr\H{o}di and S.~Schmalzbauer,
  EPJ Web Conf.\  {\bf 175} (2018) 07020
  [arXiv:1709.10487].
\bibitem{Borsanyi:2010cj}
  S.~Borsanyi, {\it et al},
  JHEP {\bf 1011} (2010) 077
  [arXiv:1007.2580].
\bibitem{Borsanyi:2011sw}
  S.~Borsanyi {\it et al},
  JHEP {\bf 1201} (2012) 138
  [arXiv:1112.4416].
\bibitem{Gavai:2004sd}
  R.~V.~Gavai and S.~Gupta,
  Phys.\ Rev.\ D {\bf 71} (2005) 114014
  [hep-lat/0412035].
\bibitem{Vovchenko:2017gkg}
  V.~Vovchenko {\it et al},
  Phys.\ Rev.\ D {\bf 97} (2018) no.11,  114030
  [arXiv:1711.01261].
\bibitem{Stephanov:2006dn}
  M.~A.~Stephanov,
  Phys.\ Rev.\ D {\bf 73} (2006) 094508
  [hep-lat/0603014].
\bibitem{Karsch:2010hm}
  F.~Karsch {\it et al},
  Phys.\ Lett.\ B {\bf 698} (2011) 256
  [arXiv:1009.5211].
\bibitem{DElia:2016jqh}
  M.~D'Elia, G.~Gagliardi and F.~Sanfilippo,
  Phys.\ Rev.\ D {\bf 95} (2017) no.9,  094503
  [arXiv:1611.08285].
\end{thebibliography}
\end{document}